\documentclass[sigconf, 9pt, nonacm]{acmart}
\usepackage{makecell}
\usepackage{multirow}
\usepackage{color}
\usepackage{enumitem}

\usepackage{listings}

\definecolor{dkgreen}{rgb}{0,0.6,0}
\definecolor{gray}{rgb}{0.5,0.5,0.5}
\definecolor{mauve}{rgb}{0.58,0,0.82}

\AtBeginDocument{%
  }

\setcopyright{acmlicensed}
\copyrightyear{2018}
\acmYear{2018}
\acmDOI{XXXXXXX.XXXXXXX}

\acmConference[Conference acronym 'XX]{Make sure to enter the correct
  conference title from your rights confirmation email}{June 03--05,
  2018}{Woodstock, NY}
\acmISBN{978-1-4503-XXXX-X/18/06}

\begin{document}

\title{MemIntelli: A Generic End-to-End Simulation Framework for Memristive Intelligent Computing}


\author{Houji Zhou}
\authornote{Both authors contributed equally to this research.}
\email{zhouhj@hust.edu.cn}
\orcid{0000-0003-1751-5265}

\affiliation{%
  \institution{School of Integrated Circuits, Huazhong University of Science and Technology}
  \city{Wuhan}
  \state{Hubei}
  \country{China}
}

\author{Ling Yang}
\authornotemark[1]
\email{d201980703@hust.edu.cn}
\orcid{0000-0003-1751-5265}
\affiliation{%
  \institution{School of Integrated Circuits, Huazhong University of Science and Technology}
  \city{Wuhan}
  \state{Hubei}
  \country{China}
}

\author{Zhiwei Zhou}
\email{zhou_zw@hust.edu.cn}
\affiliation{%
  \institution{School of Integrated Circuits, Huazhong University of Science and Technology}
  \city{Wuhan}
  \state{Hubei}
  \country{China}
}

\author{Yi Li}
\authornote{Corresponding authors}
\affiliation{%
 \institution{School of Integrated Circuits, Huazhong University of Science and Technology}
  \city{Wuhan}
  \state{Hubei}
  \country{China}
}
\email{liyi@hust.edu.cn}

\author{Xiangshui Miao}
\affiliation{%
 \institution{School of Integrated Circuits, Huazhong University of Science and Technology}
  \city{Wuhan}
  \state{Hubei}
  \country{China}
}
\email{miaoxs@hust.edu.cn}

\renewcommand{\shortauthors}{L. Yang, H. Zhou et al.}

\begin{abstract}
    Memristive in-memory computing (IMC)  has emerged as a promising solution for addressing the bottleneck in the Von Neumann architecture. However, the coupling between the circuit and algorithm in IMC makes computing reliability susceptible to non-ideal effects in devices and peripheral circuits. In this respect, efficient software-hardware co-simulation tools are highly desired to embed the device and circuit models into the algorithms. In this paper, for the first time, we proposed an end-to-end simulation framework supporting flexible variable-precision computing, named MemIntelli\footnote{\hyperlink{}{https://github.com/HUST-ISMD-Odyssey/MemIntelli}}, to realize the pre-verification of diverse intelligent applications on memristive devices. At the device and circuit level, mathematical functions are employed to abstract the devices and circuits through meticulous equivalent circuit modeling. On the architecture level, MemIntelli achieves flexible variable-precision IMC supporting integer and floating data representation with bit-slicing. Moreover, MemIntelli is compatible with NumPy and PyTorch for seamless integration with applications. To demonstrate its capabilities, diverse intelligent algorithms, such as equation solving, data clustering, wavelet transformation, and neural network training and inference, were employed to showcase the robust processing ability of MemIntelli. This research presents a comprehensive simulation tool that facilitates the co-design of the IMC system, spanning from device to application.
\end{abstract}



\received{5 May 2024}
\received[revised]{5 May 2024}
\received[accepted]{5 May 2024}

\maketitle

\section{Introduction}

    Intelligent computing, referred to as the algorithms and models involved in ML and AI\cite{RN13}, has gained widespread acceptance across various fields. However, the performance advancements of these models have been hindered by the limitations imposed by the processing speed and computing power of hardware platforms. To overcome this challenge, in-memory computing (IMC) on emerging memristive devices, such as resistive random access memory(RRAM), and phase change memory (PCM), become a promising solution, which offers efficient in-place dot product operations using the crossbar architecture\cite{RN15}. In recent years, a large of board-level and chip-level demonstrations have verified the acceleration ability in intelligent computing\cite{RN22}. However, hardware-level verifications are often a time-consuming and expensive process. Using verification tools to pre-model the intelligent systems offers a faster and more cost-effective approach to gain insights into system performance and identify potential issues in advance. 


    Existing simulation frameworks have largely explored the non-idealities of the memristive device and computing arrays, such as device variations and data quantization\cite{RN23}. Nevertheless, there remains a fundamental and noteworthy challenge in array-level computing: achieving the desired computing precision using multiple arrays. The fixed and finite status of a single array often falls short of meeting the diverse precision requirements of intelligent models. Strategies like the bit slice method\cite{RN24, RN25}, which utilizes multiple arrays to store a single matrix, have been extensively explored in the literature. Most of the current simulation models only focus on the simulation of a single crossbar array, disregarding the crucial aspect of array-level co-optimization. The interplay between input precision, weight precision, and out resolution in different arrays presents complex coupling effects. Pursuing finer-grained simulation on the precision promotion methods will be beneficial in providing accurate guidelines for real-world applications. Furthermore, at the architecture level, it is imperative to develop an ultra-flexible simulation framework that is compatible with well-established machine learning frameworks like PyTorch. This compatibility will enable mixed-precision model training and inference, empowering users to construct models with varying levels of precision according to their specific needs.

\begin{table*}[htbp]
    \centering
    \resizebox{\textwidth}{!}{
    \begin{tabular}{ccccccccc}
        \toprule
            \multicolumn{2}{l}{} & NeuroSim\cite{RN9, RN10} & PytorX\cite{RN4} & MemTorch\cite{RN8} & MINSIM\cite{RN20} & IBM aihwkit\cite{RN2} & CrossSim\cite{RN6} & This work\\
        \midrule
            \multicolumn{2}{c} {year} & 2017 & 2019 & 2020 & 2021 & 2021 & 2022 & 2024\\
            \multicolumn{2}{c} {Prog. Langu.} & \makecell[c]{Python, \\ C, C++} & Python & \makecell[c]{Python, \\ C, C++} & Python & \makecell[c]{Python, \\  C++} &  Python & Python \\
            \multicolumn{2}{c} {Device} & SRAM, NVM & RRAM & NVM & RRAM & NVM & - & NVM\\
            \hline
            \multicolumn{1}{c}{\multirow{3}{*}{framework}}  & PyTorch & \checkmark & \checkmark & \checkmark & \checkmark &  &  & \checkmark \\
            \multicolumn{1}{c}{}   & TensorFlow &  \checkmark&  &  &  & \checkmark & \checkmark &  \\
            \multicolumn{1}{c}{}   & NumPy &  &  &  &  &   & \checkmark & \checkmark \\


            \hline
            \multicolumn{1}{c}{\multirow{2}{*}{slice support}} & Dynamic slicing & &  &  &  &   & & \checkmark \\
            \multicolumn{1}{c}{} & FP support & &  &  &  &   & & \checkmark \\
            \hline
            
            \multicolumn{2}{c}{Neural network training} & \checkmark & \checkmark &  & & \checkmark & \checkmark & \checkmark \\

            \multicolumn{2}{c}{Mixed precision simulation} &  &  &  & & \checkmark &  & \checkmark \\
            
            \hline
            
            \multirow{2}{*}{Special layers} & Similarity & & & & & & & \checkmark \\
                            & Recurrent& & & \checkmark & & \checkmark & & \\
        \bottomrule
    \end{tabular}}
    \caption{Comparison of MemIntelli with current related open-source simulation frameworks}
    \label{table_comparison}
    \vspace{-0.5cm}
\end{table*}

In this paper, we introduce MemIntelli, a comprehensive simulation framework that encompasses the entire process of memristive intelligent computing pre-process verification. MemIntelli seamlessly integrates with well-established Python matrix computing frameworks, NumPy and PyTorch, ensuring compatibility at the application level. MemIntelli stands out with its exceptional flexibility in precision configuration at both the crossbar and system levels, achieved through the innovative dynamic bit-slicing method. The key features of this work can be summarized as follows:

\begin{enumerate}[leftmargin=*]
    \item MemIntelli introduces a novel variable-precision IMC scheme that supports integer and floating-precision data simulation utilizing bit-slicing. Taking non-idealities like array-level variations, parasitic effects, and data quantization into consideration, elaborate mathematical functions are meticulously modeled, ensuring a robust and accurate representation of the system.
    
    \item By constructing a hardware computing graph using PyTorch, we have successfully implemented a layer-wise precision configurable machine learning framework. The framework offers the capability to configure precision at the individual layer level, allowing for mixed-precision data training, inference, and direct model conversion. 
    
    \item To demonstrate the robust processing capabilities of MemIntelli, we conducted a series of experiments using various intelligent algorithms, including equation solving, data clustering, wavelet transformation, as well as neural network training and inference. These diverse algorithms were selected to showcase the wide range of tasks that MemIntelli can effectively handle, highlighting its generic processing ability in different intelligent applications.
    
\end{enumerate}

\vspace{-0.2cm}
\section{Background and Motivation}

\subsection{Memristive simulation frameworks}
    Early studies of nonvolatile simulation tools, like NVSim\cite{RN11} and NVmain\cite{RN12}, primarily aimed at constructing high-precision models to assess the hardware characteristics of circuits, including energy consumption, area, and latency. These frameworks lessly consider the accuracy-guided realization of the memristive applications. Subsequent studies shifted their focus towards system-level co-optimization of hardware and software, rather than solely pursuing high precision at the circuits. For instance, NeuroSim\cite{RN9, RN10} and MNSIM\cite{RN20} developed simulation tools that facilitate accuracy estimation within pre-defined hardware architectures. In fact, architecture design and accuracy benchmarking are interrelated and dependent aspects of IMC. Abstracting the memristive computing from the hardware architectures and realizing complete functional simulation becomes increasingly critical for real-world applications. Frameworks can isolate the operator-level nonidealities of VMM computation and try to provide guidelines for crossbar fabrication as well as application implementation of the crossbars.

\begin{figure}[htbp]
    \centering
    \includegraphics[scale=0.26]{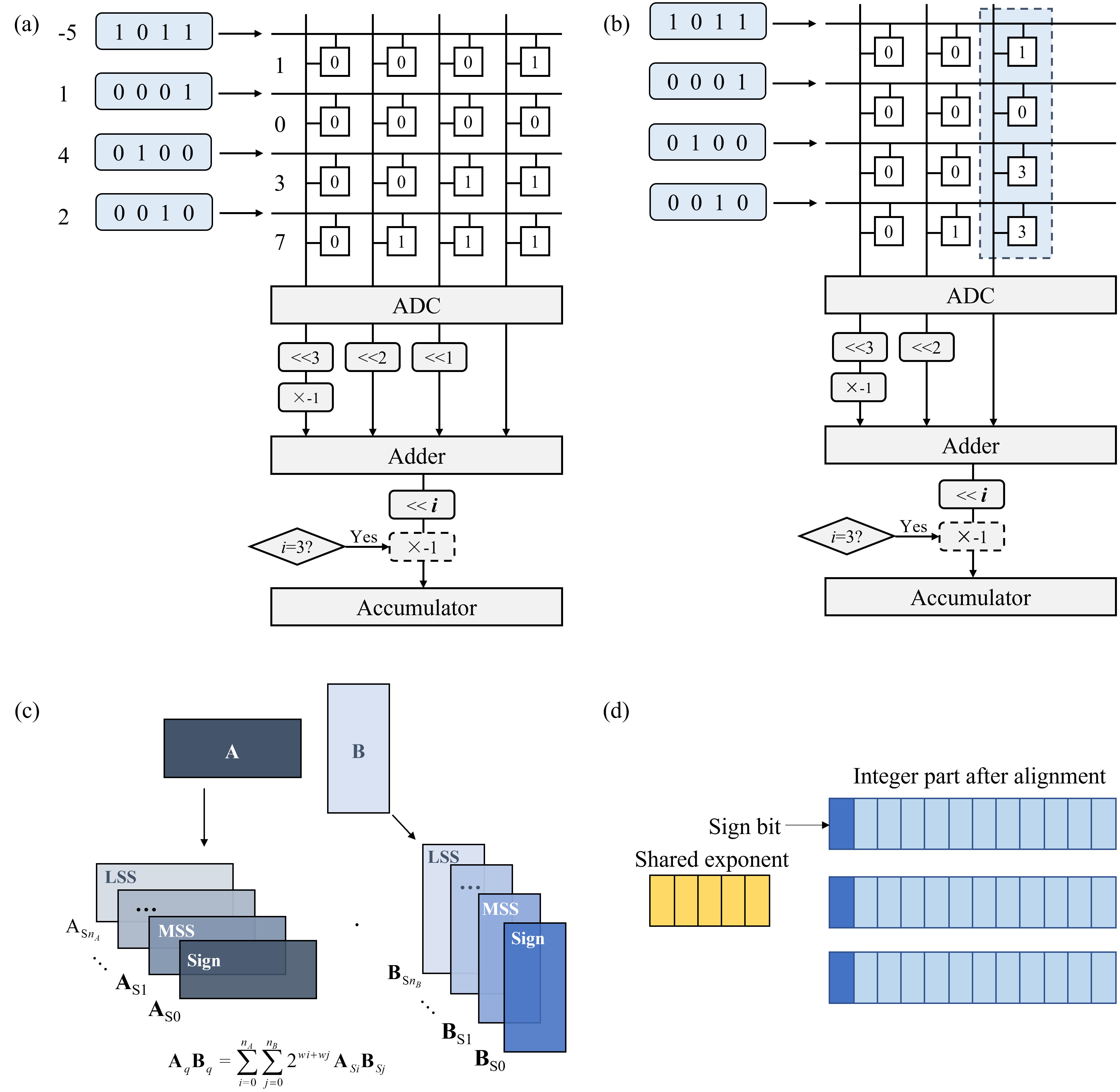}
    \caption{Mechanism of bit-slicing method. (a) Fully binary mapping. (b) Asymmetric mapping. (c) Bit-slicing matrix multiplication. (d) Shared exponent strategy for FP matrix}
    \label{bit-slicing}
    \vspace{-0.5cm}
\end{figure}

\begin{figure*}[htbp]
    \centering
    \includegraphics[scale=0.55]{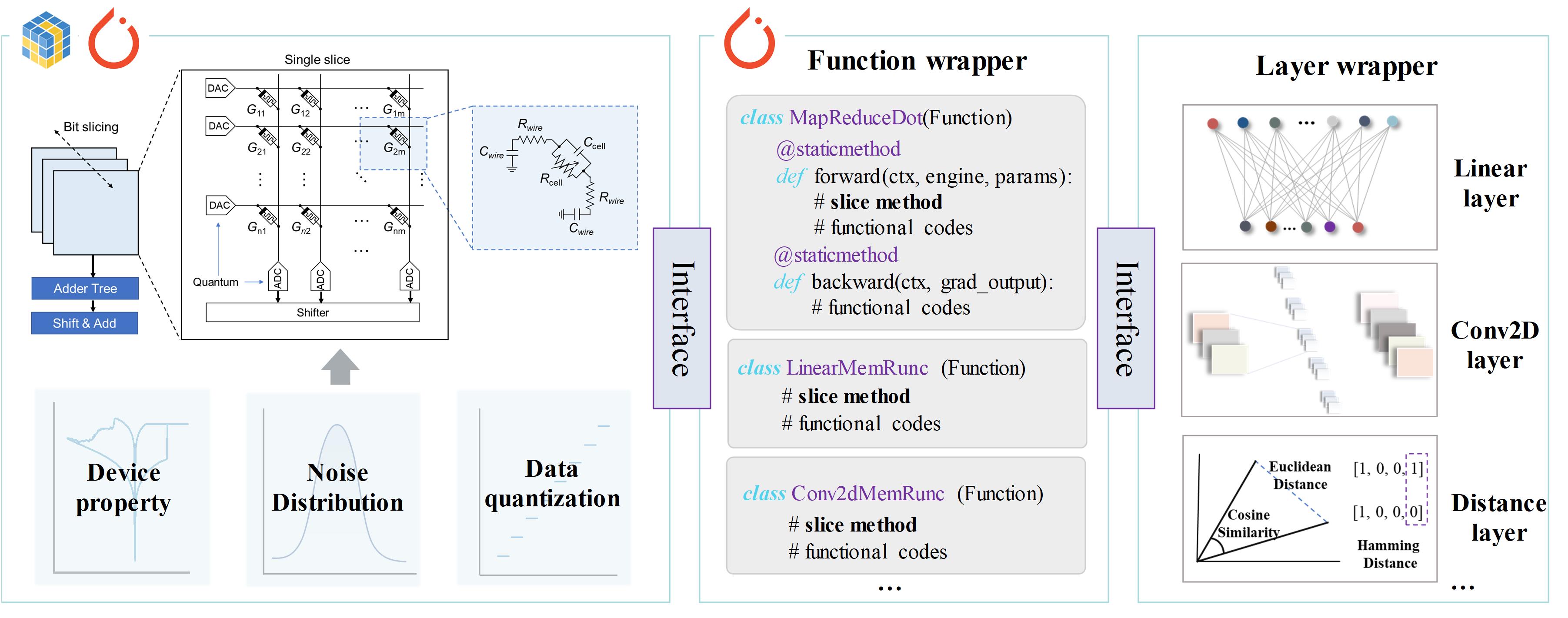}
    \caption{The overview of the MemIntelli framework}
    \label{architecture}
    \vspace{-0.3cm}
\end{figure*}

    In the realm of memristive computing, numerous simulation frameworks have been proposed to address non-idealities at the array-level modeling. One such framework, PytorX is proposed based on PyTorch and realized the crossbar-level error-aware data training\cite{RN4}. Compatibility with mature machine learning frameworks, like PyTorch and TensorFlow, become a vital feature in hardware-based simulations. Another PyTorch-based framework, MemTorch, focuses exclusively on neural network inference and incorporates more detailed device-level non-idealities. It also supports CUDA-based GPU acceleration for neural network models\cite{RN8}. Notably, the IBM research group proposed an open-source analog hardware acceleration tool kit that supports deep neural network training and inference\cite{RN2}. Their recent release includes a cloud-based simulation tool, empowering users to define hardware-aware models for training and inference using the cloud composer. Focusing on weight precision promotion, frameworks like CorssSim\cite{RN6} and configurable CNN model\cite{RN3}, can achieve evenly distributed bit-slicing storage during the inference process. These two frameworks still lack flexibility in accuracy configuration and support for network training.

\vspace{-0.2cm}
\subsection{Bit-slicing method}

    Bit-slicing represents a promising method to address the low precision issue of in-memory computing architecture\cite{le2022precision}, particularly for integer (INT) data. By distributing a number with several devices, it becomes possible to construct a high-precision IMC processor, as depicted in Figure \ref{bit-slicing}. For instance, in a simple scenario, a 4-bit signed INT could be decomposed into 4 slices, with the multiplication process being executed in four operation cycles, as illustrated in Figure \ref{bit-slicing}(a). Additionally, by leveraging the multiple conductance states of memristors, the less significant bits can be mapped to multi-bit devices, reducing the overall number of required devices, as shown in Figure \ref{bit-slicing}(b). Similarly, in the case of a matrix,  each slice could be regarded as a standalone matrix with a significance of $2^{wi}$ (see Figure \ref{bit-slicing}(c)). For floating-point (FP) data, it is hard to implement the accumulation. Herein, a shared exponent strategy is introduced to pre-alignment the FP data \cite{koster2017flexpoint}, making the elements of the matrix share the same exponent and enabling the accumulation of FP data on the crossbar array (see Figure \ref{bit-slicing}(d)).

\subsection{Motivation}
    Table \ref{table_comparison} presents a comprehensive comparison between MemIntelli and existing open-source simulation frameworks. MemIntelli distinguishes itself by highlighting two key features: variable-precision support for INT and FP data and layer-wise mixed-precision simulation for applications. Although the evenly bit-slicing method has been proposed in \cite{RN6, RN3}, it ignored the influence of the weighted slice on the most and least significant bits. Furthermore, high-precision calculation schemes for model training and scientific computing using direct FP data still lack verification solutions. Therefore, our objective in this work is to offer a flexible and variable data-slicing method that enables the realization of arbitrary precision data calculations within hardware constraints. Based on the dot product computing framework, layer-wise precision data training and inference schemes are proposed that can support the mixed-precision configuration in one model. On the application level, it seamlessly integrates with Numpy for dot product-level realization and PyTorch for machine-learning applications.

\section{MemIntelli tool design}

\subsection{MemInteli overview}
    Figure \ref{architecture} shows the overview of the MemIntelli framework. There are two main parts in MemIntelli including the dot product engine (DPE) core part and the PyTorch-compatible hardware neural network layers. At the array level, detailed circuit models that consider the non-idealities of device and functional models are established. A variable-precision method for INT and FP data is demonstrated to construct the DPE with arbitrary computing precision. For the hardware-related neural network layers, we first build the functional layers for hardware-based dot product, the linear layer, the convolutional layer, etc, for automatic computational graph construction and data backpropagation. Then based on the functional layers, the wrapped neural network layers are designed to fit the diverse usage requirements of the in-memory computing applications. The designed framework is totally compatible with the functions in PyTorch, such as the loss function, and activation.

\subsection{Crossbar module}
    In analog IMC, input voltage, memristor conductance, and output current are the parameters that represent the operands. Therefore, the stability of those three parameters determines the reliability of the computation. However, the intrinsic nonidealities including the device-to-device and cycle-to-cycle conductance variations, parasitic capacitance, IR-drop, and the noises in voltage and current are unavoidable. Analyzing the detailed influence of nonidealities on the final result is the key step in the IMC system design.
    
\begin{figure}[tbp]
    \centering
    \includegraphics[scale=0.27]{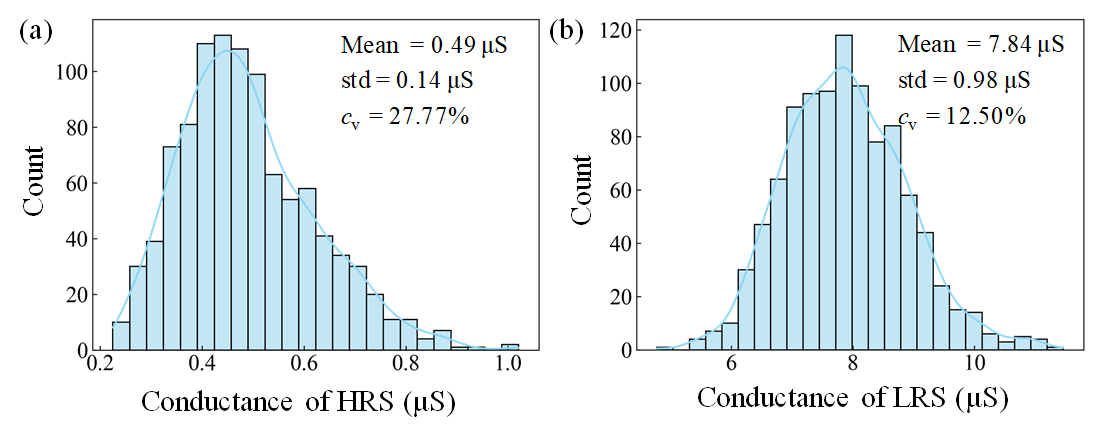}
    \caption{Conductance data generated by device model.}
    \label{conductance}
    \vspace{-0.3cm}
\end{figure}

    For the variations model of the device conductance, according to previous studies \cite{grossi2016fundamental, karpov2017log}, the statistical randomness of the memristor conductance follows the lognormal distribution. Usually, the variability of the device is described with the coefficient of variation ($c_v$) defined as $\frac{\sqrt{D(G)}}{E(G)}$, where $\sqrt{D(G)}$ is the standard deviation (std) and $E(G)$ represents the mean value of the conductance. For the lognormal function, the statistical parameters $\sigma$ and $\mu$ are given by Equation \ref{eq1}. In MemIntelli, the device-to-device and cycle-to-cycle variations are described together with the real-time random noises added to the ideal conductance matrix. According to Equation (1), using the experimental data in \cite{yang2023self}, the model can generate the simulation data that is highly consistent with the tested data in the high-resistance and low-resistance states (HRS and LRS) according to lognormal distribution, as shown in Figure \ref{conductance}. Herein, the long-term effects such as retention are not considered in the single operation. 
    
\begin{equation}
    \left\{
    \begin{aligned}
        \sigma & = \sqrt{\ln({c_v}^2+1)} \\
        \mu & = \ln(E(G))-\frac{\sigma}{2}
    \end{aligned}
    \right.
    \label{eq1}
\end{equation}

\begin{figure}[tbp]
    \centering
    \includegraphics[scale=0.33]{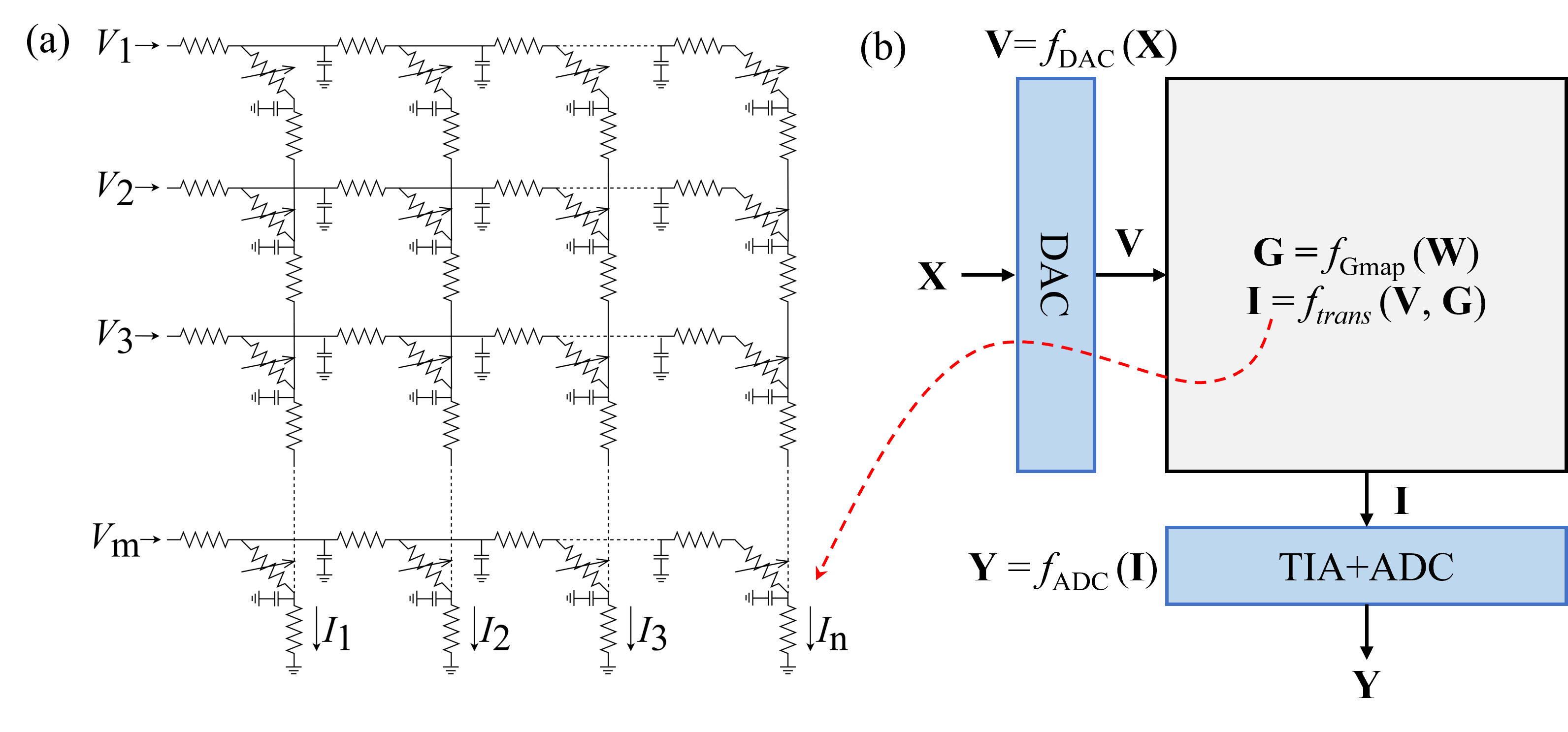}
    \caption{Simulation model of the core. (a) Circuit model of the crossbar. (b) Mathematic model.}
    \label{crossbar}
    \vspace{-0.5cm}
\end{figure}

Based on the device model, a circuit model of the crossbar array is proposed containing the memristors, wire resistances, and the parasitic capacitance of the device and connected wires, as shown in Figure \ref{crossbar}(a). In this model, the relationship between the output current and the input voltage, as well as the conductance matrix, is considerably more intricate than the dot product based on Ohm's law and Kirchhoff's law of current. Differential equations are developed to accurately represent this relationship. Additionally, it is important to note that the digital-to-analog and analog-to-digital data converters (DAC and ADC) introduce quantization errors during the computing processes. To facilitate simulation, all mappings and conversions are mathematically described, encompassing the DAC, ADC, conductance mapping, and the signal transmission model within the crossbar array as shown in Figure \ref{crossbar}(b). 

\subsection{Variable-precision hardware DPE module}

\begin{figure}[tbp]
    \centering
    \includegraphics[scale=0.3]{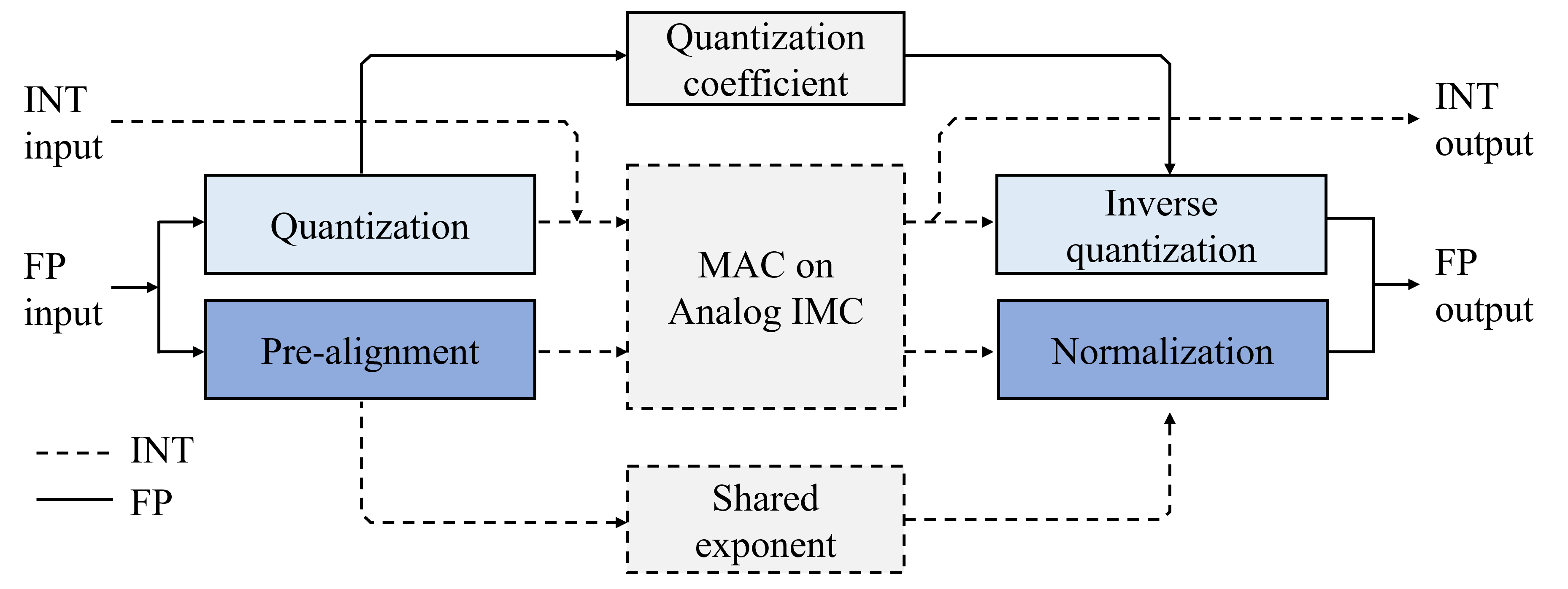}
    \caption{The matrix multiplication flow of INT data and FP data.}
    \label{core}
    \vspace{-0.2cm}
\end{figure}

Dot product operation is the fundamental function of the IMC architecture. In this work, a variable-precision hardware DPE module is proposed based on the bit-slicing mechanism. In this architecture, INT data could be processed directly, while all the FP input will be converted to INT format before implementing the computing by quantization or pre-alignment with the quantization coefficients and shared exponent stored in the register to recover the FP results finally. Therefore, the INT and FP data could share the same IMC core circuit, as shown in Figure \ref{core}. 

\begin{figure}[tbp]
    \centering
    \includegraphics[scale=0.3]{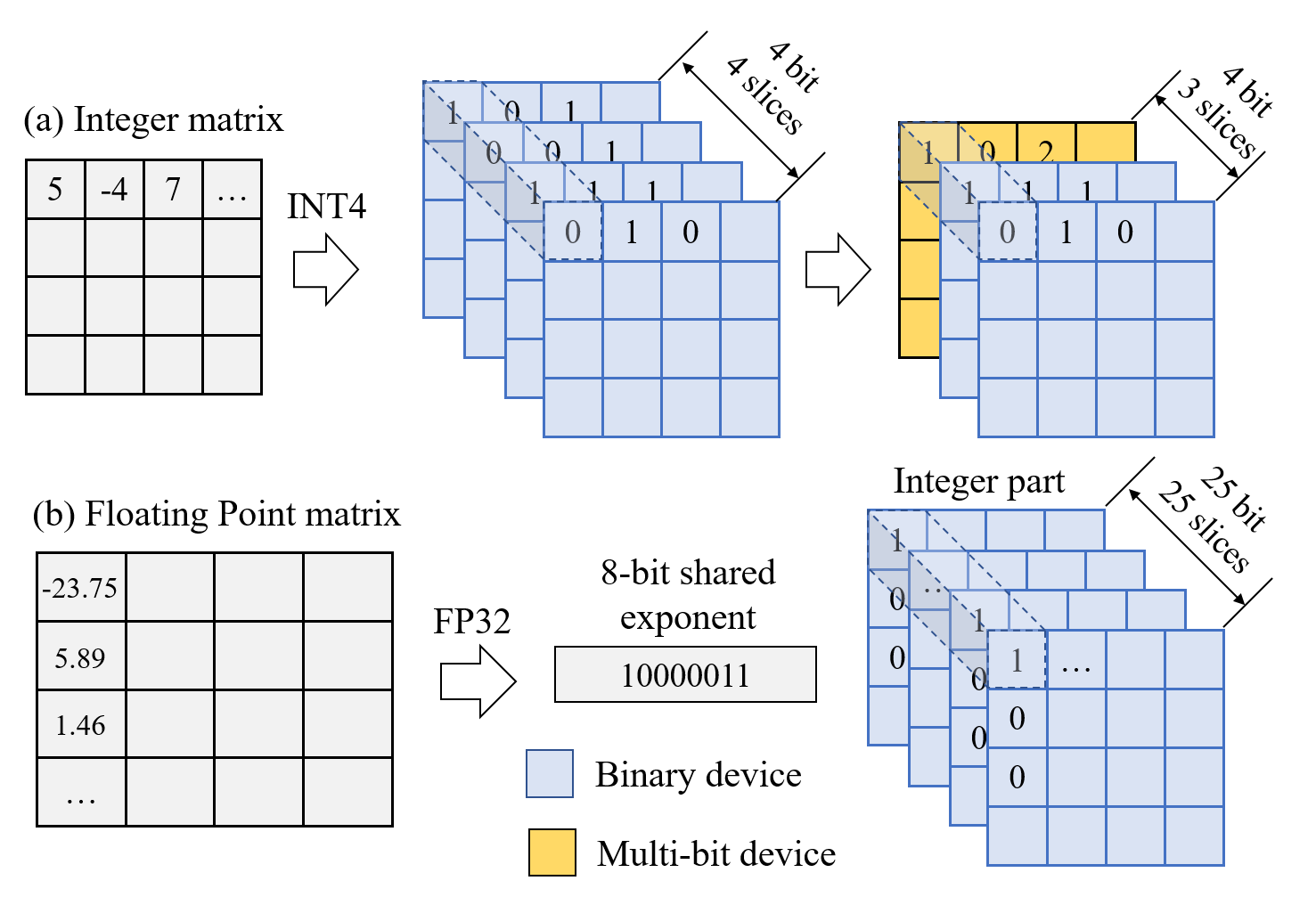}
    \caption{Variable-precision in-memory computing. Two examples in the cases of (a) INT4 and (b) FP32 matrix.}
    \label{variable-precision}
    \vspace{-0.3cm}
\end{figure}

To realize variable-precision computing flexibly, the different slices of the original matrix are mapped onto the different array groups, where each group consists of several arrays. In this case, the hardware could adapt the effective bit widths of the operands flexibly by activating a corresponding number of array groups. For example, to perform the multiplication for an N-bit-format matrix decomposed into $N_s$ slices ($N_g \geq N_s$) with an $N_g$-group system, it requires activating $N_s$ array groups for the storage and computing for the $N_s$ slices, as shown in Figure \ref{variable-precision}. 

\begin{figure}[tbp]
    \centering
    \includegraphics[scale=0.42]{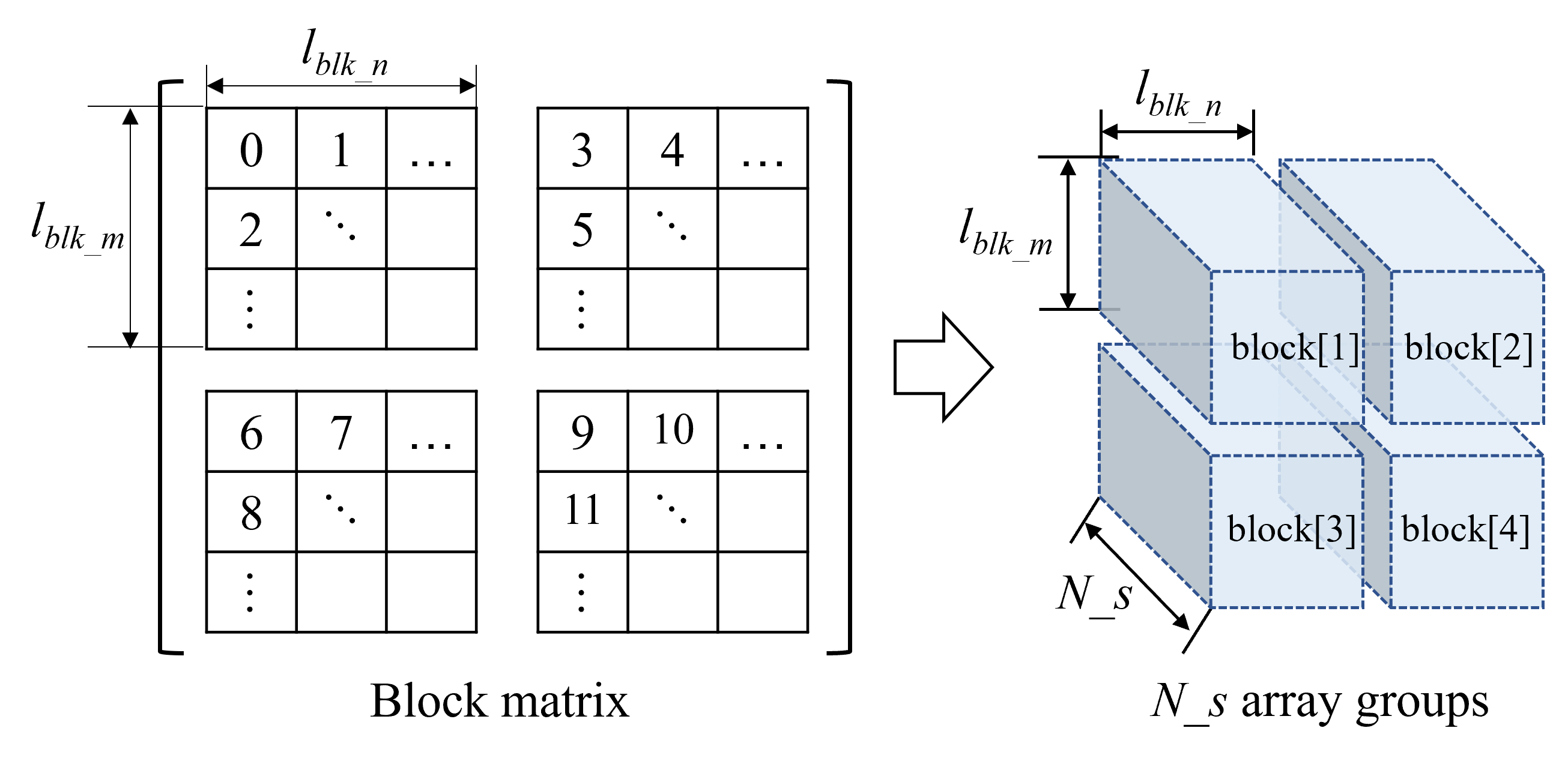}
    \caption{Block matrix mapping.}
    \label{blk_matrix}
    \vspace{-0.5cm}
\end{figure}

Besides, since the size of the matrices to be mapped and the size of the arrays are often different, the original matrix is decomposed into several submatrices in terms of the size of the array $l_{blk\_m} \times l_{blk\_n}$, as shown in Figure \ref{blk_matrix}. The errors caused by quantization and pre-alignment usually increase with the matrix size. Herein, based on the block matrix mapping strategy, quantization and pre-alignment are also performed in terms of the block size to reduce the errors from preprocessing by scaling down the size of the matrix to be quantized or pre-aligned. In this case, the elements within the same blocks share a quantization coefficient or an exponent. If the matrix size is not divisible by the array size, it is padded with zeros to make up the difference.

\subsection{Hardware neural network module with computing graph}

\begin{figure}[tbp]
    \centering
    \includegraphics[scale=0.47]{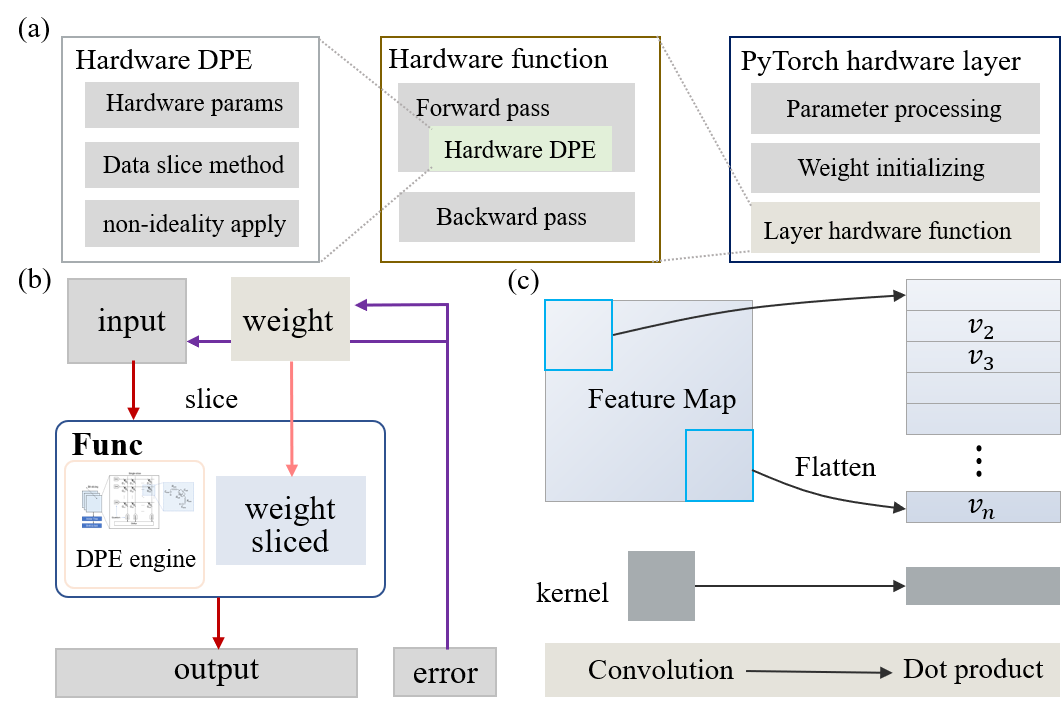}
    \caption{Design principles of PyTorch comparable hardware layers. (a)The developing concept. (b) Dataflow in hardware function realization. (c) The img2col algorithm concept.}
    \label{fig4}
    \vspace{-0.5cm}
\end{figure}

Figure \ref{fig4}(a) demonstrates the principles of constructing the hardware computing layers using PyTorch. Based on the hardware DPE, MemIntelli constructed the computing function with a computing graph that supports the automatic forward and back-propagation pass. Further, the hardware layers, including the linear, convolutional layers, and similarity layers, are realized. 

\textbf{Hardware function with computing graph.} Figure \ref{fig4}(b) shows the computing process when realizing the hardware computing layers in PyTorch. According to the configuration of the DPE, the hardware function receives the full precision input and weight data. For the forward pass, the function will process the input and weight data to the customized quantization methods to obtain the forward outputs. In practice, to reduce repeated calculations in the weight quantization process, the quantization data of the weights will be saved as an attribute in the computing graph and passed to the computing function directly. During the backward pass, the errors are directly applied to the full precision weight and input data to ensure the model is trainable and not trapped in the local minimum as most work declared.

In the functional realization, only dot product operation can be realized by the engine. For the convolution, the operational matrix can not be mapped to the crossbar arrays directly. For example, the 2D convolution often requires three-dimensional input feature maps and convolution kernels. These high-dimensional matrices need to be flattened and converted to 2D arrays when required for the implementation on the crossbar array hardware. Figure \ref{fig4}(c) shows an example of converting the 2D convolution to the dot product process using the image-to-column (img2col) algorithm \cite{wu2023review}. Then the forward pass of the convolution is converted to the dot product and then applied to the hardware dot engines.


\textbf{PyTorch hardware layer setup.} The Hardware layer wraps the hardware function and realizes a high-level description for the intelligent model construction. The following codes show the example of parameters set up in the linear layer (namely \textit{LineraMem}). In each layer, the users must provide the input and weight slice methods (referred to as the parameters \textit{input\_sli\_med} and \textit{weight\_sli\_med}) for the layer initialization. The slice methods support the INT and FP data as introduced earlier. During the forward pass, these slice methods are used in the hardware computing graph for the dot products. Additionally, the DPE is dependent on the layer setup, which provides much flexibility for the configuration. As mentioned before, the hardware layers realized the weight quantization and kept the sliced copy of the weight. Compared with the layer models in \textit{torch.nn}, the hardware layer fully inherited the parameters while adding the hardware parameters. Thus, models trained by full-precision data can be directly transferred to the hardware ones just using \textit{torch.load\_state\_dict}. Then call the function of \textit{update\_weight()}, and the parameters will converted to the hardware-wise ones.

\begin{lstlisting}
class LinearMem(nn.Module):
    def __init__(self, engine, in_features, out_features, input_sli_med, weight_sli_med, bias, device=None):
        # functional code
    def forward(self, x):
        # functional code
    def update_weight():
        # functional code
\end{lstlisting}

\begin{figure}[tbp]
    \centering
    \includegraphics[scale=0.6]{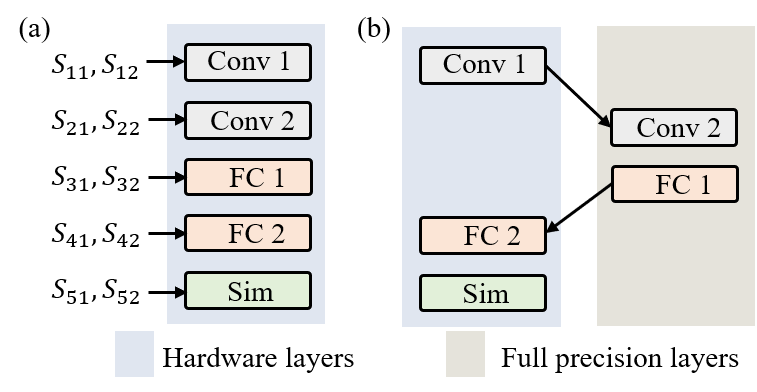}
    \caption{The examples of the reconfigurability of the hardware layers. (a) The layer-wise data slicing ability. (b) The mixed hardware layers and full-precision layers structure.}
    \label{configurebility_layers}
    \vspace{-0.3cm}
\end{figure}

\textbf{Ultra flexible layer-wise configuration.} 
Benefiting from the dependent hardware engine and the extremely high model compatibility, MemIntelli can support ultra-flexible simulation degrees of freedom for the model construction. Figure \ref{configurebility_layers} shows two typical configured structures in the model. Figure \ref{configurebility_layers}(a) demonstrates that the slice methods in each layer are locally aware of the model configuration and will not affect the other layers. Then, MemIntelli can support the different inter-layer precision simulations for the design of space exploration like \cite{yuan2021nas4rram}. Additionally, heterogeneous compute engines are even supported on different hardware layer implementations. Figure \ref{configurebility_layers}(b) shows the ability of the hybrid system simulation where some precision-sensitive layers can be deployed on the digital system. The partly hardware layers are used for pursuing efficiency. These solutions improve the diversity in hardware space exploration.

\begin{figure}[tbp]
    \centering
    \includegraphics[scale=0.25]{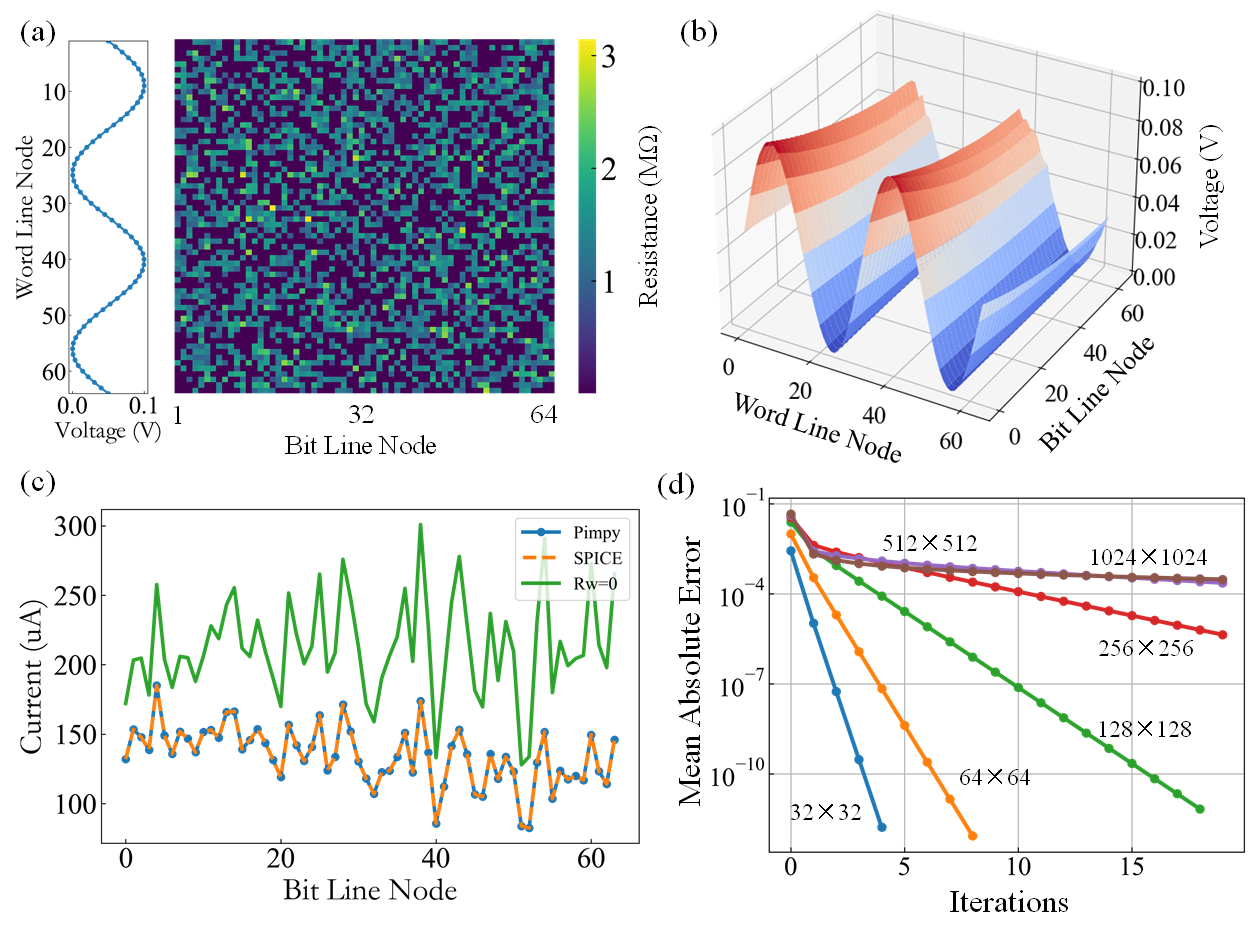}
    \caption{Simulation results of the single array with the wire resistance of $2.93\   \Omega$. (a) The tested $64\times64$ random conductance matrix and the input voltage vector applied on the word lines. (b) Voltage attenuation on the word line caused by IR-drop. (c) Output currents on the bit lines in different cases. (d) Model solving performance in different array sizes.}
    \label{model_test}
    \vspace{-0.3cm}
\end{figure}

\section{DPE core Verification}
Based on the circuit model of the crossbar array shown in Figure \ref{crossbar}, MemIntelli provides an efficient tool for physical simulation. To verify the proposed model, a $64 \times 64$ scale array with a wire resistance of $2.93\ \Omega$ and a discrete sinusoidal input voltage sequence at the word line was used for testing, as shown in Figure \ref{model_test}(a). Due to the IR-drop caused by the wire resistance, there is a voltage attenuation on the word line as shown in Figure \ref{model_test}(b), resulting in the decrease of the output currents (Figure \ref{model_test}(c)). The simulation result of our model is identical to that obtained by LTspice software. Besides, a cross-iteration algorithm is proposed to accelerate the simulation speed of the crossbar model, even for the $1024 \times 1024$ array, the model can generate a solution with the error below $10^{-3}$ within 20 iterations, as shown in Figure \ref{model_test}(d).

\begin{figure}[tbp]
    \centering
    \includegraphics[scale=0.27]{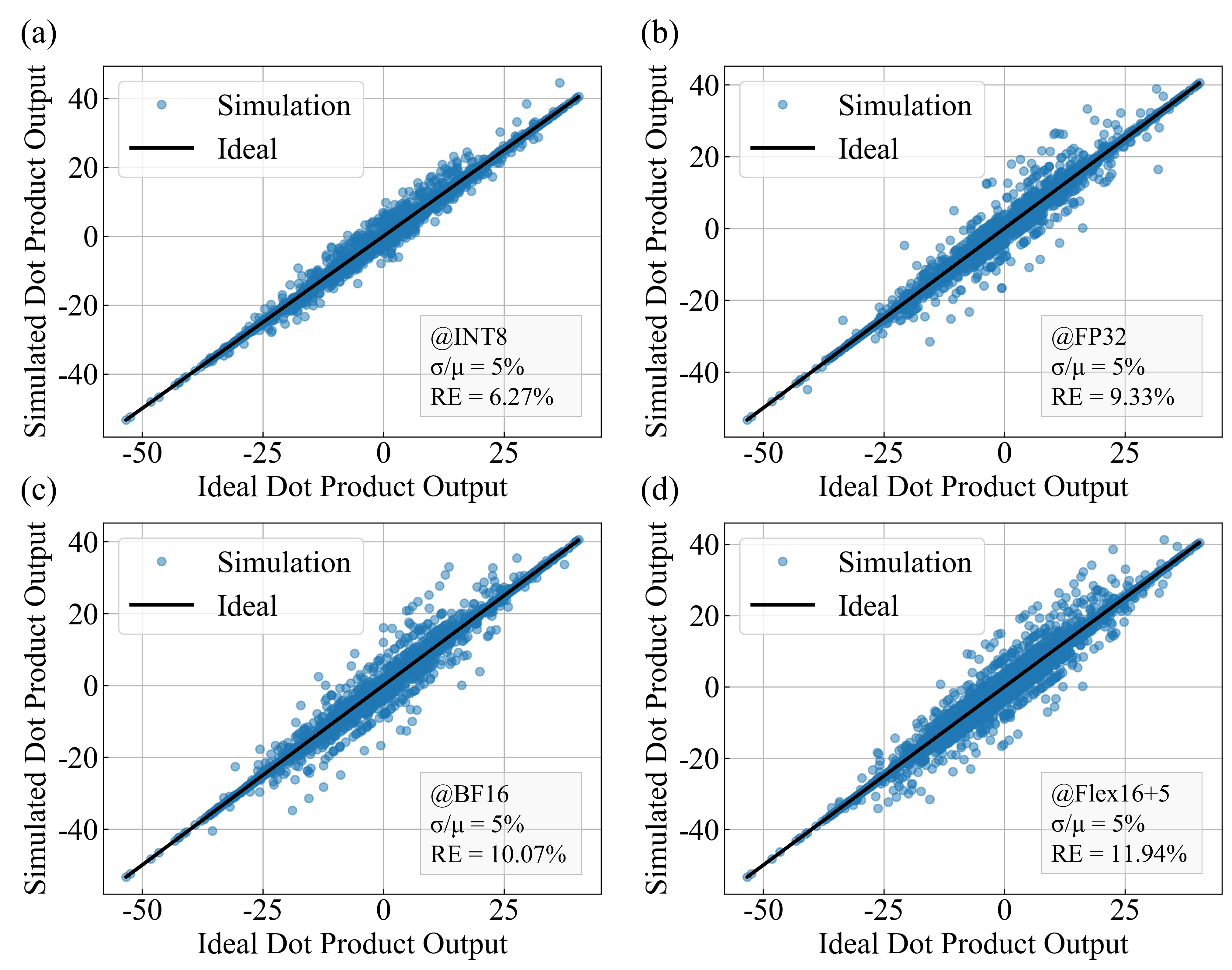}
    \caption{Variable-precision matrix multiplication simulation results with the matrix size of $128 \times 128$. (a) INT8. (b) FP32. (c) BF16. (d) FlexPoint16+5.}
    \label{vmm_test}
    \vspace{-0.2cm}
\end{figure}

To verify the variable-precision matrix multiplication function, $\mathbf{A} \cdot \mathbf{B} = \mathbf{C}$ is tested, where the sizes of the random matrix $\mathbf{A}$ and $\mathbf{B}$ are $128 \times 128$ and the original data format of FP64. Figure \ref{vmm_test} depicts the simulation results in different data formats. The relative error (RE) of the dot product is computed as $\frac{\|{Simulation-Ideal}\|_2}{\| Ideal \|_2}$.

\begin{figure}[tbp]
    \centering
    \includegraphics[scale=0.31]{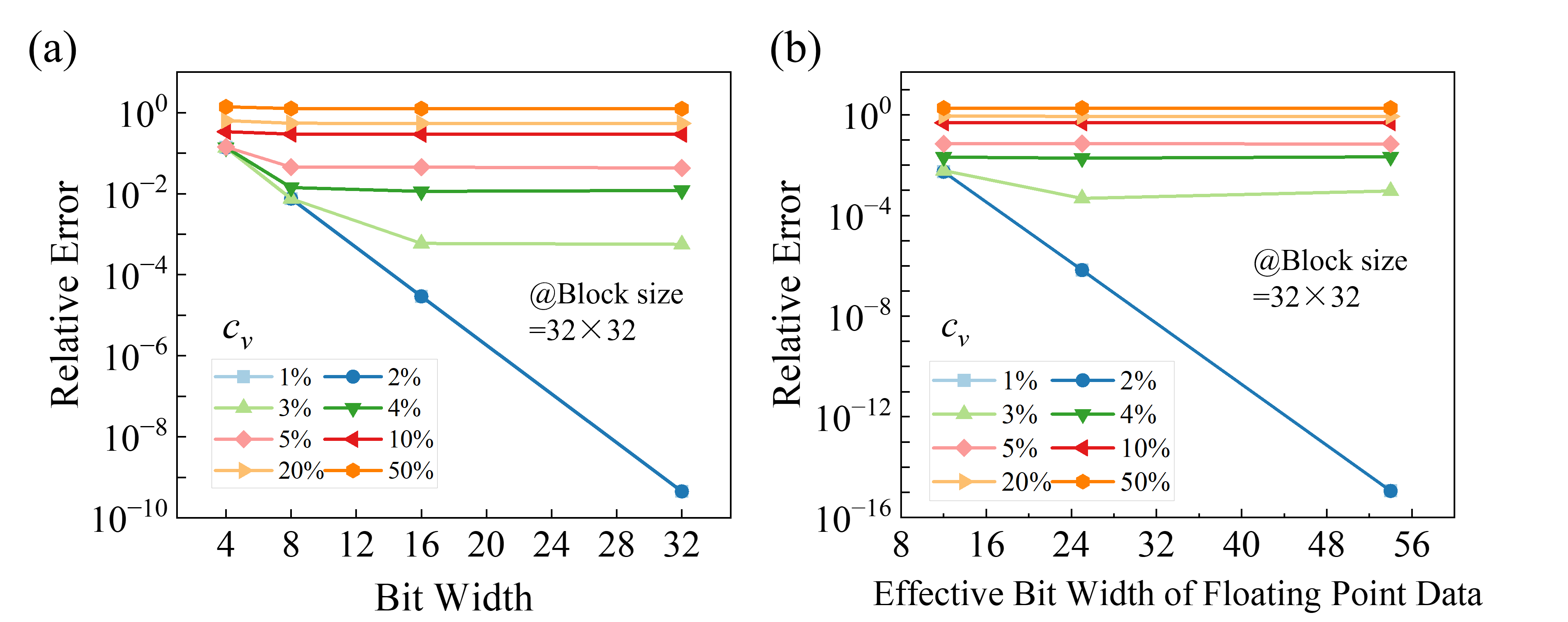}
    \caption{Monte Carlo simulation for 100 cycles in the given conditions to analyze the influence of the nonidealities. (a) Quantization: conductance variations and block size. (b) Pre-alignment: conductance variations and block size.}
    \label{mc}
\end{figure}

Monte Carlo simulation is one of the most important methods in circuit designing. MemIntelli also provides an efficient Monte Carlo analysis tool for the nonidealities such as conductance variation, quantization, and size of the array. Figure \ref{mc} shows the influence of the bit width, block size, and conductance variations. This result indicates that it is hard for IMC to achieve software accuracy without using any calibration operation. Besides, it is found that the quantization-based dot product could achieve a lower relative error compared with the pre-alignment method in the same hardware parameters and effective bit width, where the effective bit width denotes the length of the INT part after the pre-alignment.

\section{EXPERIMENTAL RESULTS}
In order to verify the generic proficiency of MemIntelli in processing a wide range of intelligent algorithms, we shall present a selection of application illustrations in which memristive computing has played a pivotal role \cite{bring_zhou}. These exemplary instances shed light on various abilities of MemIntelli, encompassing precision prerequisites, model inference, training, and more. Parameters used in the simulation are shown in table \ref{params} if not specified.

\begin{table}[tbp]
    \centering
    \setlength{\tabcolsep}{1.2mm}{
    \begin{tabular}{cccccccc}
         \toprule
         Params & HGS &  LGS & g\_levels & var & rdac & radc & array\_size \\
         \midrule
         Values & $ 1e-5 $ & $ 1e-7 $ & 16 &  0.05 & 256 & 1024 & (64, 64) \\
         \bottomrule
    \end{tabular}}
    \caption{Experimental parameters used for the verification in MemIntelli.}
    \label{params}
    \vspace{-0.5cm}
\end{table}

\begin{figure}[tbp]
    \centering
    \includegraphics[scale=0.26]{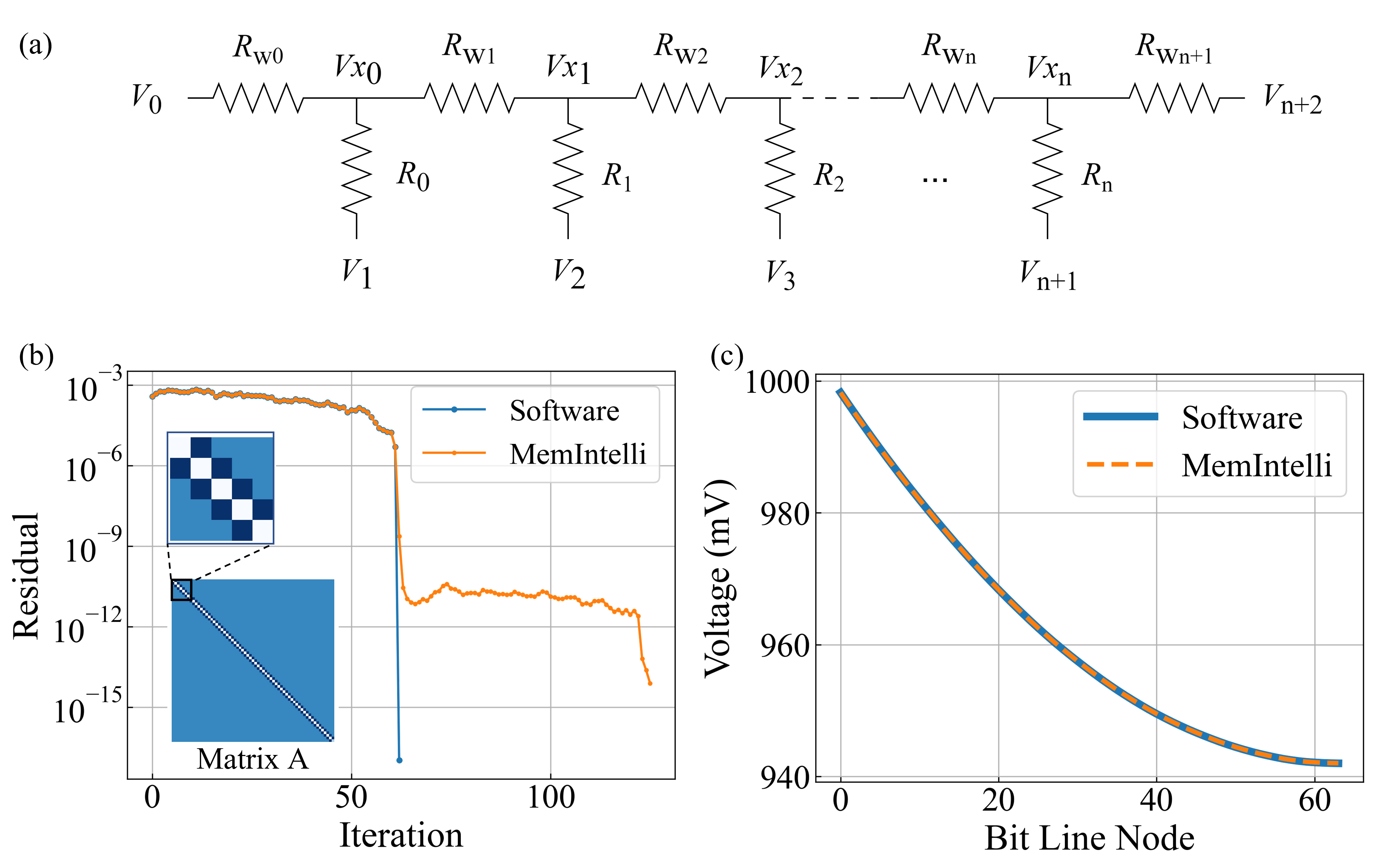}
    \caption{Linear equation solving task. (a) The equivalent circuit of the word line is modeled using the linear equation. (b) Solving process comparison with conjugate gradient algorithm. (c) Results of the circuit equation solving.}
    \label{Eq_solve}
    \vspace{-0.5cm}
\end{figure}

\textbf{Solving memristive circuit equation.} Linear equation solving is the core of many scientific and engineering computing tasks, involving iterative dot product operations. IMC could serve as a highly efficient hardware solver for linear equations utilizing the parallelism DPE. As a case study, we present the solving of the circuit equation for the word line with wire resistance, as shown in Figure \ref{Eq_solve}(a). To calculate the node voltages $[V_{x0}, V_{x1}, ..., V_{xn}]$, the circuit is modeled by a band linear equation according to Ohm's law and Kirchhoff's circuit laws as shown in the subgraph of Figure \ref{Eq_solve}(b). The coefficient matrix \textbf{A} is mapped on the memristive array with pre-alignment FP32 format. The simulation parameters are depicted in Table 2 but the block size is $32\times32$. Herein, the conjugate gradient algorithm is used to solve the band equation. As shown in Figure \ref{Eq_solve}(b), the software solver achieves convergence fast after several iterations. Due to the noise from hardware and pre-alignment, the convergence speed of the hardware solver is reduced in the high-precision region, therefore, it requires more iterations to achieve software-comparable precision. However, such precision is enough for fundamental circuit verification as shown in Figure \ref{Eq_solve}(c). The solving results of the software and hardware are highly consistent.

\begin{figure}[tbp]
    \centering
    \setlength{\abovecaptionskip}{0.cm}
    \includegraphics[scale=0.26]{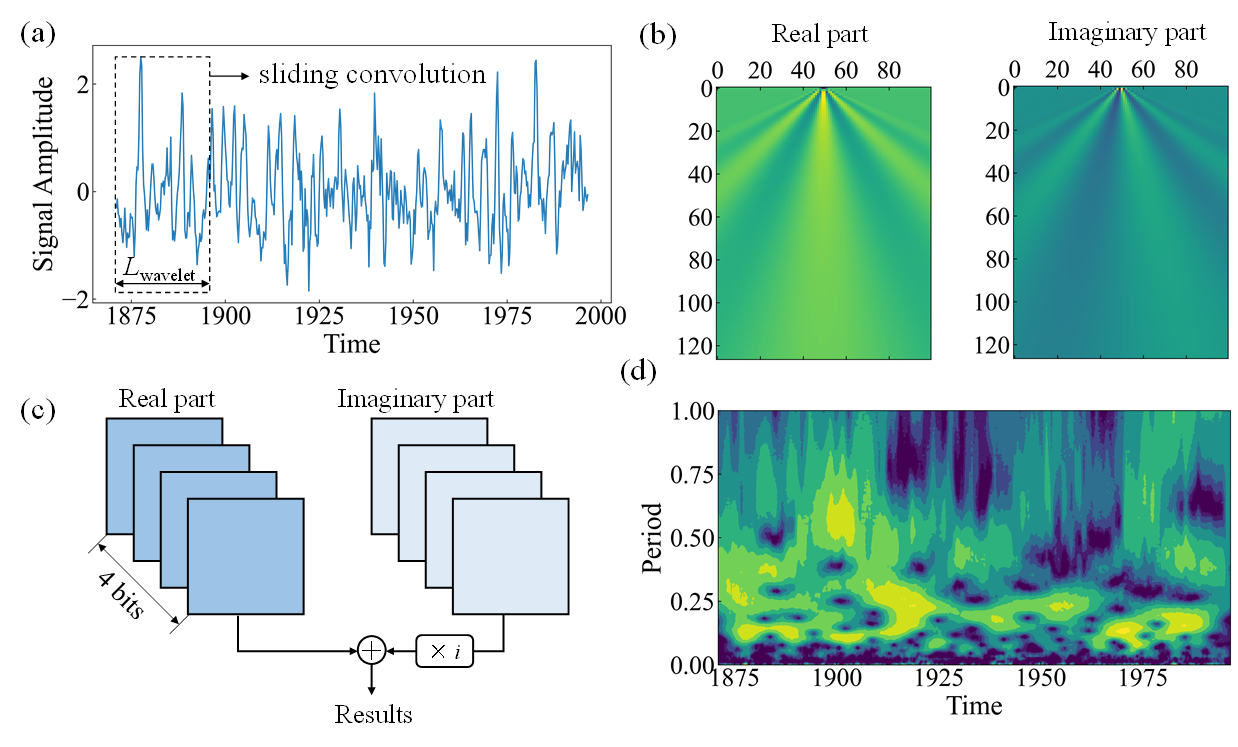}
    \caption{Continuous wavelet transform based on the Morlet wavelet. (a) Original time-serial signal from the dataset NINO3. (b) The real part and imaginary part of the Morlet wavelet complex matrix. (c) Mapping the real part and imaginary part, respectively by quantizing to INT4. (d) Power spectrum of the signal.}
    \label{cwt}
    \vspace{-0.5cm}
\end{figure}

\textbf{Signal processing: continuous wavelet transform (CWT).} CWT is one of the most important signal processing techniques, especially in the field of time-series and dynamic signal processing. However, CWT involves massive sliding convolutional operations, requiring huge computing resources. In this work, by organizing the kernels as a matrix, the convolution operations are realized by matrix multiplication. Figure \ref{cwt}(a) shows the time-series signal from oel-Nino dataset \cite{misc_el_nino_122}. Herein, CWT is performed with Morlet wavelet whose kernels matrix is shown in Figure \ref{cwt}(b). Morlet wavelet is a complex wave. Therefore, the real and imaginary parts of the matrix are mapped separately by quantizing to signed INT4 format, as shown in Figure \ref{cwt}(c). The signal is convolved using the real and imaginary matrices respectively and final power spectrum is obtained by integrating the convolution results of the real and imaginary parts as shown in Figure \ref{cwt}(d). 

\begin{figure}[htbp]
    \centering
    \includegraphics[scale=0.38]{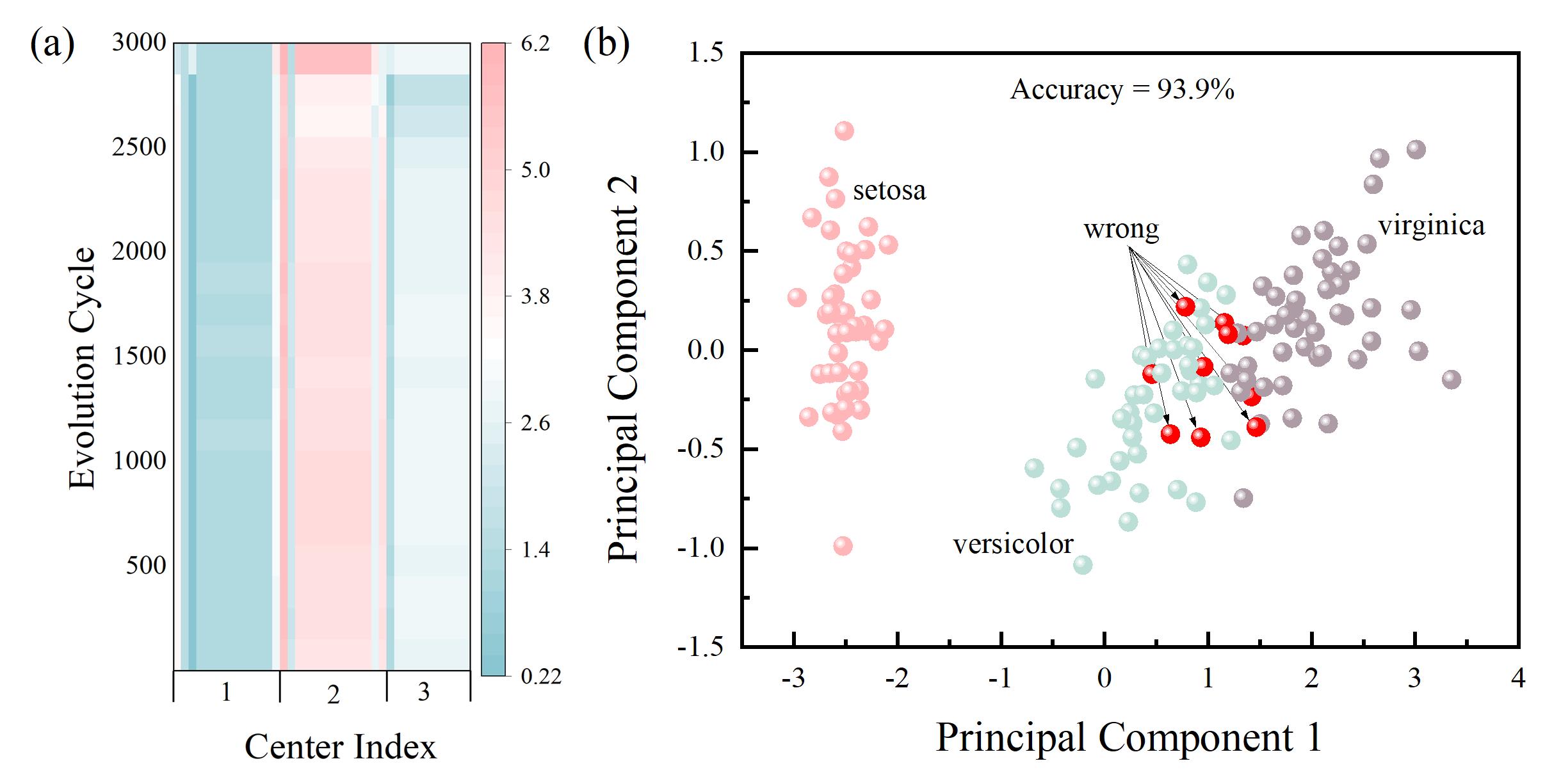}
    \caption{Experimental verification on data clustering tasks. (a) Weight evolution of the centers (b) Data clustering results using the IRIS dataset.}
    \label{fig2}
    \vspace{-0.3cm}
\end{figure}

\textbf{Data clustering.} Data clustering task involves largely similarity measurement operations, such as the Euclidean distance and cosine similarity. These operations have largely relied on the dot product operations and have been studied extensively \cite{RN30,jeong2018k}. Here, we show an example of realizing the K-means data clustering using Euclidean distance. Since the Euclidean distance calculation involves the squared items that are not supportive of the crossbar array, several approximate calculation schemes have been proposed. Here, we follow the realization of \cite{RN30}, and the Euclidean distance is calculated by $ (x-y)^2 \approx -2x\cdot y_i + y_i^2$. Here x is the input vector and $ y_i $ is the center in the K-means algorithm. Then using the hashing method, the Euclidean distance is converted to the dot product results of $ [x, -1/2, -1/2, ..., -1/2] $ and $ [y, y^2/n, y^2/n, ..., y^2/n] $ where n pieces of -1/2 and $ y^2/n $ are spliced at the tail of x and y respectively. Here, we adopt the IRIS dataset \cite{misc_iris_53} as the toy demonstration and n here is 10. The precision of the data is set to INT8 with the slice method of (1, 1, 2, 4) and only one center is updated during the iteration. Figure \ref{fig2}(a) shows the iterated data in MemIntelli and the clustering results are shown in \ref{fig2}(b) where the clustering results are counterparts with the full precision computing.

\textbf{Neural network training.}
Constrained by the precision of the NVM devices for the weight storage, many neural networks implemented on the NVM array only concentrate on the data inference which maps the well-trained weights on the crossbars. However, data training on analog hardware is much more important to promote the accuracy of the implemented models where the models can learn to be more robust to resist the non-idealities of the hardware. Here, we utilize Memintelli to realize the data training process of  LeNet-5 using the MNIST dataset\cite{RN29}. We demonstrated the training results in precision of three types of data formats: INT4 (slice to 1, 1, 2), INT8 (sliced to 1, 1, 2, 4), and FP16 (sliced to 1, 1, 2, 4, 4 ) to evaluate the performance. 

\begin{figure}[htbp]
    \centering
    \includegraphics[scale=0.38]{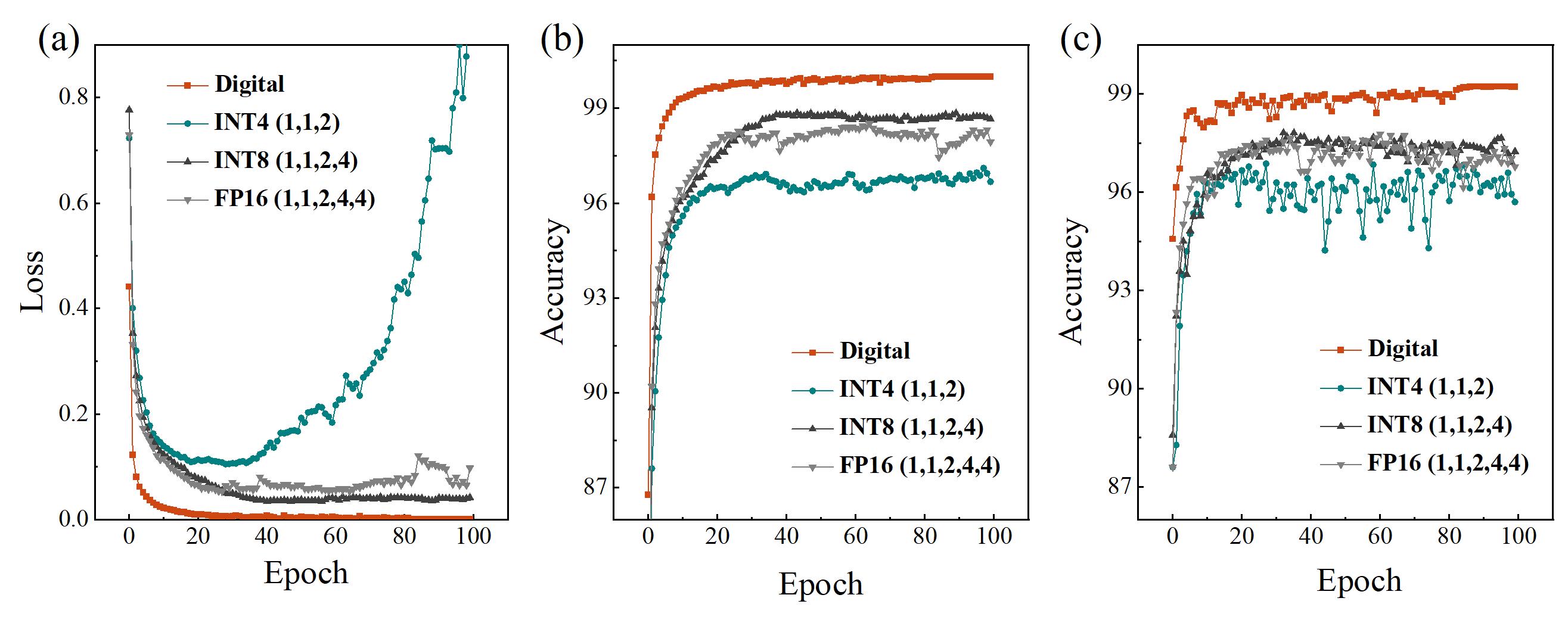}
    \caption{Experimental verification on training the LeNet-5 under the data formats INT4, INT8, and FP16. (a) The training loss. (b)The training accuracy. (c) The test accuracy.}
    \label{fig6}
    \vspace{-0.3cm}
\end{figure}

Figure \ref{fig6} shows the training results of LeNet-5 on MemIntelli. In general, the limited precision training shows different degrees of accuracy loss compared to the full precision one.  Over-quantized data (weight and input) may lead the model not to converge, for example, INT4 precision in figure \ref{fig6}(a). Additionally, INT precision may have a higher effective bit width than the FP data as mentioned in figure \ref{mc}, and have higher accuracy for both training and inference.

\begin{figure}[htbp]
    \centering
    \includegraphics[scale=0.38]{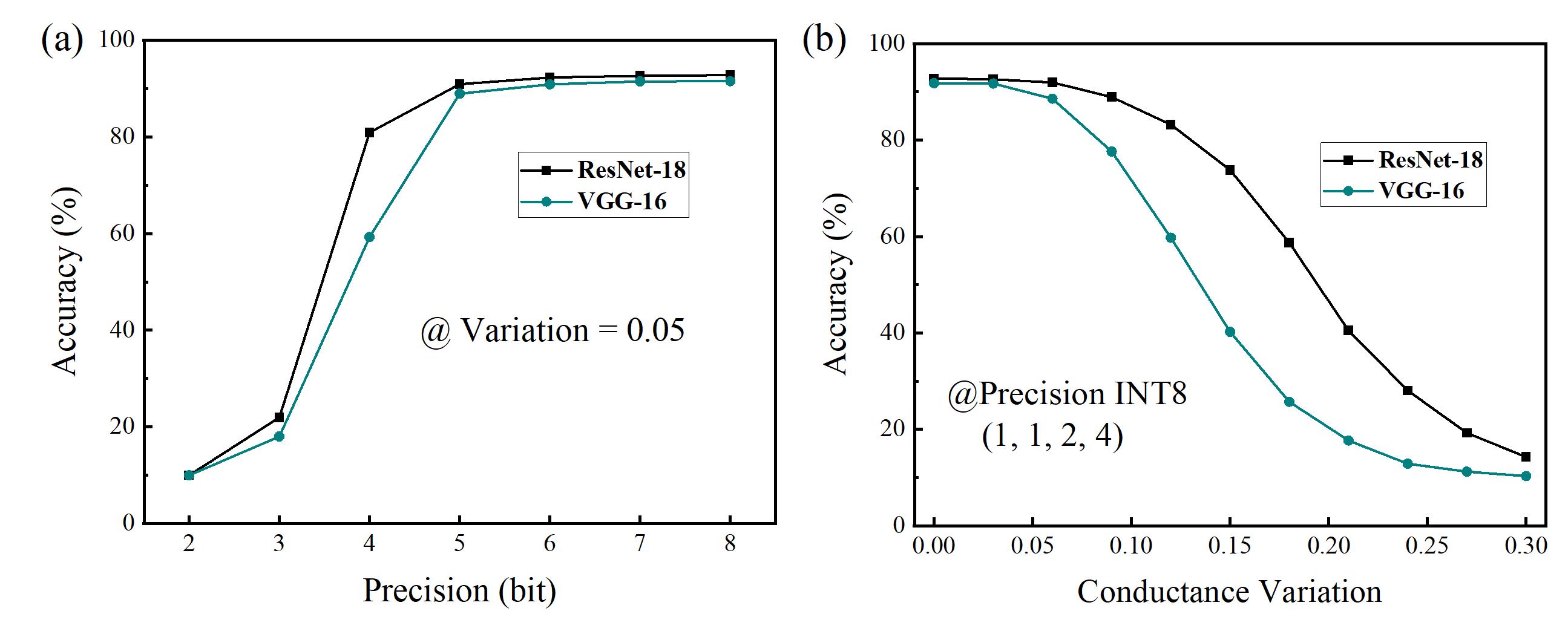}
    \caption{ResNet-18 and VGG-16 experimental inference verification using MemIntelli. (a) The inference results under different quantization bits. (b) The accuracy under different levels of noise of a single array.}
    \label{fig5}
    \vspace{-0.3cm}
\end{figure}

\textbf{Neural network inference}.
MemIntelli supports the data directly mapping from the well-trained models to the hardware platform without additional model converting and modifying. Here, ResNet-18\cite{RN26} and VGG-16\cite{RN27} neural networks are employed to verify the inference process on MemIntelli. Figure \ref{fig5} shows the simulation results of the networks using the CIFAR-10 dataset. We first demonstrate the sliced pieces of weight and input influencing the results. to simplify the configuration, the input and weight data share the same slice method. All the data are sliced to one bit, for example, INT4 is sliced to (1, 1, 1, 1). The experimental results are shown in figure \ref{fig5} (a). Results show that the accuracies of the network have large losses under the slice bits under 5. The accuracy larger than 5 bits reaches a plateau period where the loss can be lower than 3\%. Figure \ref{fig5} (b) demonstrates the influence of the variation of a single array on the accuracy of the models. Variation larger than 5\% will make the accuracy decrease dramatically.

Table \ref{table_runtime} shows the runtime performance of realizing the NN models using MemIntelli on the GPU and CPU platforms. Here we benchmarked the slice methods of FP16 where slices are (1, 1, 2, 4, 4). On CPU realization, the total amount of the number of the parameters affects the speed greatly where larger models require more computing time. However, in the GPU version, VGG-16 shows a higher processing speed than ResNet-18. This is due to VGG net has more intensive matrix operations than ResNet which can make full use of the computing resources of the GPU.

\begin{table}[htbp]
    \centering
    \resizebox{\linewidth}{!}{
    \begin{tabular}{cccccc}
        \toprule
        Datasets & \makecell[c]{Image \\size} & \makecell[c]{Batch \\size} & Model & \makecell[c]{CPU \\ runtime} & \makecell[c]{GPU \\ runtime}\\
        \midrule
        CIFAR-10 & 3*32*32 & 128 & ResNet18 & 0.27 img/s & 7 img/s\\
        CIFAR-10 & 3*32*32 & 128 & VGG-16 & 0.143 img/s& 16 img/s\\
        MNIST & 1*28*28 & 128 & LeNet5 & 166.4 img/s & 1920 img/s\\
        \bottomrule
    \end{tabular}}
    \caption{The performance of MemIntelli when applied on the neural network inference. The GPU uses the Nvidia GeForce RTX 3080 Ti and CPU uses the 12th Gen Intel i7-12700, ~2.1GHz.}
    \label{table_runtime}
    \vspace{-0.5cm}
\end{table}

\section{Conclusion}
In conclusion, this work demonstrated MemIntelli, an end-to-end simulation framework, that supports generic memristive intelligent computing algorithms realization. Elaborate models are established on the array and circuit level to construct the memristive crossbar simulation. A variable-precision method for INT and FP data is proposed based on the bit-slicing method to model the precision-adjustable vector computing cores. On the architecture level, MemIntelli can realize layer-wise configurable mixed precision simulation. In the future, we plan to add more complex device models, such as the conductance drift, on the array to pursue a more realistic simulation process. Additionally, extra functional layers, including recurrent and transformers, and design space exploration tools will be completed in future versions.


\begin{acks}
We acknowledge financial support from the STI 2030—Major Projects (2021ZD0201201), the National Key Research and Development Plan of MOST of China (2019YFB2205100, 2022YFB4500101), the National Natural Science Foundation of China (92064012), and the Fundamental Research Funds for the Central Universities, HUST:5003190006. The Hubei Engineering Research Center on Microelectronics and Chua Memristor Institute also supported this work.
\end{acks}

\newpage
\bibliographystyle{ACM-Reference-Format}
\bibliography{sample-base}

\end{document}